\documentclass[
 reprint,
 superscriptaddress,
 amsmath,amssymb,
 aps,
]{revtex4-2}

\usepackage{graphicx}
\usepackage{comment}
\usepackage{float}
\usepackage[caption=false]{subfig}
\usepackage{dcolumn}
\usepackage{bm}
\usepackage[table]{xcolor} 
\usepackage{tabularx}
\usepackage{booktabs}
\usepackage{multirow}
\usepackage{array}    
\usepackage[
    colorlinks=true,
    linkcolor=blue,    
    citecolor=blue,    
    urlcolor=black      
]{hyperref}

\begin{document}

\preprint{APS/123-QED}

\title{Gravitational Wave Signal Denoising and Merger Time Prediction By Deep Neural Network}

\author{Yuxiang Xu}
\email{xuyuxiang22@mails.ucas.ac.cn}
\affiliation{Center for Gravitational Wave Experiment, National Microgravity Laboratory, Institute of Mechanics, Chinese Academy of Sciences, Beijing 100190, China}
\affiliation{Hangzhou Institute for Advanced Study, UCAS, Hangzhou 310024, China}
\affiliation{Shanghai Institute of Optics and Fine Mechanics, Chinese Academy of Sciences, Shanghai 201800, China}
\affiliation{Taiji Laboratory for Gravitational Wave Universe (Beijing/Hangzhou), University of Chinese Academy of Sciences (UCAS), Beijing 100049, China}
\author{He Wang}
\email{hewang@ucas.ac.cn}
\affiliation{Taiji Laboratory for Gravitational Wave Universe (Beijing/Hangzhou), University of Chinese Academy of Sciences (UCAS), Beijing 100049, China}
\affiliation{International Centre for Theoretical Physics Asia-Pacific, University of Chinese Academy of Sciences, 100190 Beijing, China}
\author{Minghui Du}
\affiliation{Center for Gravitational Wave Experiment, National Microgravity Laboratory, Institute of Mechanics, Chinese Academy of Sciences, Beijing 100190, China}
\author{Bo Liang}
\affiliation{Center for Gravitational Wave Experiment, National Microgravity Laboratory, Institute of Mechanics, Chinese Academy of Sciences, Beijing 100190, China}
\affiliation{Hangzhou Institute for Advanced Study, UCAS, Hangzhou 310024, China}
\affiliation{Shanghai Institute of Optics and Fine Mechanics, Chinese Academy of Sciences, Shanghai 201800, China}
\affiliation{Taiji Laboratory for Gravitational Wave Universe (Beijing/Hangzhou), University of Chinese Academy of Sciences (UCAS), Beijing 100049, China}
\author{Peng Xu}
\email{xupeng@imech.ac.cn}
\affiliation{Center for Gravitational Wave Experiment, National Microgravity Laboratory, Institute of Mechanics, Chinese Academy of Sciences, Beijing 100190, China}
\affiliation{Hangzhou Institute for Advanced Study, UCAS, Hangzhou 310024, China}
\affiliation{Taiji Laboratory for Gravitational Wave Universe (Beijing/Hangzhou), University of Chinese Academy of Sciences (UCAS), Beijing 100049, China}
\affiliation{Lanzhou Center of Theoretical Physics, Lanzhou University, Lanzhou 730000, China}

\date{\today}

\begin{abstract}
The mergers of massive black hole binaries could generate rich electromagnetic emissions, which allow us to probe the environments surrounding these massive black holes and gain deeper insights into the high energy astrophysics. However, due to the short timescale of binary mergers, it is crucial to predict the time of the merger in advance to devise detailed observational plans. The overwhelming noise and the slow accumulation of signal-to-noise ratio in the inspiral phase make this task particularly challenging. To address this issue, we propose a novel deep neural denoising network in this study, capable of denoising a 30-day inspiral phase signal. Following the denoising process, we perform the detection and merger time prediction based on the denoised signals. Our results demonstrate that for a 30-day inspiral phase data with a signal-to-noise ratio between 10 and 50 occurring no more than 10 days before the merger, our absolute prediction error for the merger time is generally within 24 hours.

\end{abstract}

\maketitle

\section{\label{sec:introduction}INTRODUCTION}

In 2015, Advanced LIGO and Advanced Virgo achieved the first direct detection of gravitational waves (GW) \cite{abbott2016gw150914}, marking the beginning of a new era in astronomy. Led by LIGO-Virgo collaboration, nearly a hundred GW events have been discovered \cite{abbott2016observation,abbott2016gw151226,abbott2016binary,scientific2017gw170104,abbott2017gw170608,abbott2017gw170814,abbott2017gw170817,abbott2020gw190425,abbott2020gw190412,abbott2020gw190521,abbott2021tests,ezquiaga2021hearing,bozzola2021general,abbott2019gwtc,abbott2021gwtc,abbott2023gwtc}, greatly enriching our understanding of the universe. However, due to terrestrial noise, ground-based GW detectors typically exhibit sensitivity to frequencies above 10 Hz \cite{abbott2019gwtc,abbott2020prospects,freise2010interferometer}, restricting the scope of astronomical exploration. Consequently, initiatives have been taken to establish space-based observation systems, such as the Laser Interferometer Space Antenna (LISA) \cite{amaro2017laser,baker2019laser}, Taiji \cite{gong2011scientific,hu2017taiji}, and Tianqin \cite{luo2016tianqin}, to circumvent the limitations and broaden the observable frequency range. Space-based GW detectors are expected to be able to detect GW signals in the range of 0.1mHz - 1Hz. Within this frequency range, a diverse array of astrophysical sources is anticipated to be detected, including massive black hole binaries (MBHBs), stellar-mass black hole binaries, extreme mass-ratio inspirals, galactic double white dwarfs, and the stochastic gravitational wave background. These detections are expected to provide novel insights and enhance the understanding of cosmic phenomena.


Among these signals, coalescing MBHBs, with signal-to-noise ratios (SNRs) ranging from tens to several thousand \cite{amaro2017laser,barack2004lisa,mangiagli2020observing}, are considered the most promising GW sources detectable and thus are designated as the primary science goal for space-based GW detection missions. The merger of MBHBs is anticipated to along with a high black hole accretion rate and a high star formation rate in galaxy \cite{di2005energy,springel2005modelling}, which could potentially result in strong electromagnetic (EM) emissions. The EM emission from MBHBs is expected to manifest in various forms during the inspiral \cite{tang2018late,d2018electromagnetic}, merger \cite{armitage2002accretion,merritt2002tracing,palenzuela2009binary}, and ring-down phases \cite{schnittman2008infrared}. Observations of EM emission provide the possibility to gain insights into the environment surrounding massive black holes. In recent years, substantial research has concentrated on GW and EM multi-messenger observations of MBHBs \cite{mangiagli2022massive,mangiagli2020observing,ivezic2019lsst,pratten2023precision}. These observations are poised to enhance our understanding of the co-evolution of massive black holes, nuclear star clusters, and their host galaxies \cite{inayoshi2020assembly}. Moreover, the concurrent analysis of GW and EM emission will further our comprehension of the evolution of galaxies at high redshifts \cite{cornish2007search}. The feasibility of this multimessenger observation largely depends on the ability of space-based GW detectors to accurately localize the source, as well as the capability of telescopes to point to the source's location. A typical binary black hole system can be localized with a median precision of $\sim 10^2$ deg$^2$ ($\sim 1$ deg$^2$) at 1 month (1 hour) from the merger \cite{mangiagli2020observing}. Candidate EM telescopes, including the Square Kilometre Array \cite{dewdney2009square} in radio wave, the Large Synoptic Survey Telescope \cite{abell2009lsst} in optical band, and Athena \cite{nandra2013hot} in X-ray, may require localization precisions ranging from 0.1 to 10 deg$^2$. It is expected that for some sources, multimessenger observations of GW and EM emission may be enabled within a few hours to a few days.

However, the observable duration of some EM emissions may only last for several hours or even shorter periods\cite{palenzuela2010dual,paschalidis2021minidisk,bogdanovic2022electromagnetic,bode2011mergers}, presenting new challenges for multi-messenger observations of mergers of MBHBs. Consequently, it is necessary to conduct pre-merger detection, i.e., predicting the time of the merger, in order to devise detailed and timely observation plans. This enhances the feasibility of achieving multi-messenger observations for mergers of MBHBs. In principle, the matched filtering algorithm, which is the predominant method for detecting GW signals, yields the most precise results \cite{finn1992detection,usman2016pycbc,cannon2021gstlal}. However, to obtain the template that most closely matches the observed GW signal, an extensive template bank is required \cite{davies2020extending,PhysRevD.90.082004,sachdev2020early}. In addition, the computational inefficiency of the matched filtering algorithm poses a significant challenge. Cornish et al. \cite{PhysRevD.105.044007} introduced a low-latency detection pipeline based on matched filtering that can analyze several months of LISA data within a few hours. Additionally, for Tianqin data, Chen et al. \cite{chen2024near} designed a search algorithm based on matched filtering by reducing the number of frequency points computed and selecting simpler likelihood functions for probability assessment. However, this also requires approximately 10 hours to execute the search. 
Given that a large number of sources are likely to be detected in the future \cite{mangiagli2022massive}, it is essential to develop faster and more accurate search algorithms to further reduce computational complexity and resource consumption.

Deep learning algorithms have demonstrated remarkable efficacy across various GW data analysis tasks due to their robust feature representation capabilities and rapid inference proficiency. These applications include signal detection \cite{george2018deep,gabbard2018matching,wang2020gravitational,krastev2020real,lopez2021deep}, parameter estimation \cite{gabbard2022bayesian,dax2021real}, and signal extraction \cite{torres2016denoising,wei2020gravitational,shen2019denoising,chatterjee2021extraction,xu2024gravitational,wang2024waveformer}. Specifically, for the pre-merger detection of MBHB signals, Ruan et al. recently proposed a deep learning model, which enables the detection of GW signals several hours to days before the merger \cite{ruan2024premerger}. This development provides the possibility of the simultaneous GW and EM detection of MBHBs. However, this model sets a stringent detection threshold of 0.999 for identifying GW signals. While this threshold significantly reduces false positives, it markedly decreases the model's sensitivity to GW signals. This is because they trained the model using noisy data directly. However, the SNR of the inspiral phase signal is too low, and the GW features in the data without denoising are too weak, making it difficult for the model to distinguish between pure noise and GW signals. Therefore, further denoising is necessary to improve the distinguishability between noise and GW signals. Additionally, the model does not provide the estimations of the time when the mergers take place, which will be the key parameters for the planning for the detailed planning of observations for EM emissions. The difficulty of pre-merger detection largely stems from the way how the SNR of GW signals from coalescing MBHB are accumulated with time \cite{mangiagli2020observing}. Especially, the signal from the final hour before the merger contains 99\% of the total SNR, causing signals from the inspiral phase to be submerged in overwhelming noise. Consequently, achieving a detectable SNR during the inspiral phase requires substantial time accumulation \cite{amaro2017laser}, and the increased length of the signal significantly burdens the training of deep learning models \cite{schafer2023first}. Based on the issues discussed above, in this work, we design a deep learning-driven workflow for denoising, detecting, and predicting the corresponding merger time of inspiral signals. The core component of this workflow is WaveUformer-CNN. WaveUformer is utilized to denoise 30-day data from the inspiral phase, while the CNN is tasked with predicting the time of the merger events. The design of WaveUformer integrates the effective dimensionality reduction capabilities of WaveUnet \cite{stoller2018wave} with the noise separation properties of Sepformer \cite{subakan2021attention}, enabling it to rapidly and efficiently denoise GW signals. Considering the enhanced prominence of features in the denoised signals, the prediction network is intentionally simplistic to prevent overfitting, comprising only a simple CNN architecture.

Our model is capable of denoising 30 days of inspiral phase signals with a SNR between 10 and 50, and predicting the corresponding merger time within 0.01 seconds. Our tests encompass signal denoising, signal detection, and merger time prediction. For the denoising test, our model demonstrated excellent denoising performance, effectively recovering the amplitude and phase of the signals. Specifically, 91.98\% of the denoised data achieved an overlap $\geq$ 0.99 with the template data, and 95.82\% of the denoised data has an SNR loss $\leq$ 5\% compared to the template SNR. For signal detection, we explored the optimal thresholds for detecting signals using our workflow. The results indicate that our workflow efficiently avoids false positives from pure noise while maintaining high sensitivity to GW signals. Our workflow achieved a average detection accuracy of 99.27\% for GW signals. Regarding merger time prediction, the model's prediction error decreases as the merger event approaches, and data with larger chirp masses tend to yield smaller errors. The absolute error in the merger time prediction for 89.1\% of the data is $\leq$ 24 hours.

The organization of this paper is as follows. In Sec~\ref{sec:datasets}, we introduce the data generation process. In Sec~\ref{sec:method}, we present the workflow used in this work, along with the core deep learning model architecture. Sec~\ref{sec:strategy} describes the training strategy of the deep learning model. The results obtained from testing the proposed method on the test set are presented in Sec~\ref{sec:results}. Finally, Sec~\ref{sec:summary} provides a summary.

\section{\label{sec:datasets}Data preparations}

In this work, the noisy signal $x(t)$ serves as the input for the denoising network, while the GW signal $h(t)$ is used as the corresponding label. For the merger time prediction network, the output of the denoising network $\hat{x}(t)$ is utilized as the input, and the merger time $t_{\text{c}}$, is used as the corresponding label. It is important to note that the above data undergoes whitening preprocessing before being used.

The noisy signal \( x(t) \) can be expressed by the equation:
\begin{equation}
x(t) = h(t) + n(t) \,,
\end{equation}
where \( n(t) \) represents the noise from the detector. The preparation of this noise follows the simulation process described below. To address the laser frequency instability noise inherent in space-based GW detection missions, Time-Delay Interferometry (TDI) \cite{Tinto,armstrong1999time,Babak2021} is commonly employed as a pre-processing technique to suppress this noise. Consequently, the noise floor in the output data from TDI channels is primarily determined by the combined effects of residual acceleration noise $n_{ACC}$ from the Gravitational Reference Sensors (GRS) and optical readout noise $n_{OMS}$. The overall noise power spectral density (PSD) is a combination of these two noise components, and the specific combination can be found in Ref~\cite{Babak2021,Tinto,wang2020numerical,wang2023revisiting}. In this work, we have considered only the optimized TDI channels \( A \) and \( E \). Based on the designs of Taiji \cite{luo2020brief} we assume the nominal values
\begin{equation}
S_{\text{OMS}}^{1/2}(f) = 8 \times 10^{-12} \sqrt{1 + \left(\frac{2\text{mHz}}{f}\right)^4} \, \frac{\rm m}{\sqrt{\rm Hz}} \,,\label{eq:SOMS}
\end{equation}
\begin{equation}
\begin{split}
S_{\text{ACC}}^{1/2}(f) = & 3 \times 10^{-15} \sqrt{1 + \left(\frac{0.4\text{mHz}}{f}\right)^2} \\
& \times \sqrt{1 + \left(\frac{f}{8\text{mHz}}\right)^4} \, \frac{\rm m/s^2}{\rm \sqrt{Hz}}\,.
\label{eq:SACC}
\end{split}
\end{equation}
We will utilize this PSD combination to obtain simulated noise that complies with the Taiji mission. At the same time, this PSD will also be used for whitening the GW signal.

We have selected the bbhx \cite{katz2020gpu} package to generate template waveform $h(t)$ with a duration of one year using the IMRPhenomD \cite{husa2016frequency,khan2016frequency} model. bbhx enables the rapid generation of a large number of frequency-domain MBHB waveforms using GPU acceleration. Importantly, all waveforms generated by bbhx have already undergone TDI processing, and we have adjusted the relevant parameters to conform to the Taiji mission specifications. The parameter space for the template data is shown in Tab~\ref{tab:data}.
\begin{table}[htb]
\caption{\label{tab:data}Summary of parameter setups in MBHB GW signals generations.}
\begin{ruledtabular}
\renewcommand{\arraystretch}{1.2}
\begin{tabular}{lcc}
Parameter & Lower bound & Upper bound\\
\colrule
$M_{\text{c}}$ \footnote{Chirp mass of the system.} & \( 10^5 M_{\odot} \) & \( 10^7 M_{\odot} \) \\
$q$ \footnote{Mass ratio of the objects in the binary system.} & 0.1 & 1 \\
$s_1^z$ \footnote{Dimensionless spin parameter for the primary object.} & -0.99 & 0.99 \\
$s_2^z$ \footnote{Dimensionless spin parameter for the secondary object.} & -0.99 & 0.99 \\
$inc$ \footnote{Inclination angle of the orbital plane with respect to the line of sight.} & 0 & \(\pi\) \\
$beta$ \footnote{Latitude angle on the sky.} & \(-\frac{\pi}{2}\) & \(\frac{\pi}{2}\) \\
$lam$ \footnote{Longitude angle on the sky.} & 0 & \(2\pi\) \\
$psi$ \footnote{Polarization angle of the GW.} & 0 & \(\pi\) \\
$t_{\text{c}}$ \footnote{Moment of the merger event within one year.} & 0.11 & 1.0 \\
\end{tabular}
\end{ruledtabular}
\end{table}
We select data spanning 30 days within 0 to 10 days preceding the merger. The sampling frequency of the data is 0.1 Hz, so each data will have 259,200 data points. The template is injected with a specific optimal SNR:
\begin{equation}
SNR = (h | h)^{1/2},
\label{eq:snr}
\end{equation}
where the inner product \( (h | \hat{x}) \) is defined as,
\begin{equation}
(h | \hat{x}) = 2 \int_{f_{\text{min}}}^{f_{\text{max}}} \left(\tilde{h}(f) \tilde{\hat{x}}^{*}(f) + \tilde{h}^{*}(f) \tilde{\hat{x}}(f)\right) \, df \,.
\label{eq:inner_product}
\end{equation}
Here, \( f_{\text{min}} = 3 \times 10^{-5} \) Hz and \( f_{\text{max}} = 0.05 \) Hz. The tilde symbol represents the Fourier transform, and \( * \) the complex conjugation. The overlap between the waveform \( \hat{x} \), extracted by the model from the noise data, and the whitened template waveform \( h \), can be evaluated using the overlap function:
\begin{equation}
\mathcal{O}(h, \hat{x}) = \max_{t_c, \phi_c} (h|\hat{x}) \,,
\end{equation}
where \( t_c \) and \( \phi_c \) are the instantaneous time and phase corresponding to the maximum overlap between  \( h \) and \( \hat{x} \).

The chosen range for the luminosity distance is from \(5 \times 10^5\) to \(5 \times 10^6\) Mpc, based on the ldc2a \cite{baghi2022lisa}. We further analyzed the SNR of the 30-day inspiral signal under this setup, and the results show that the SNR can range from $\sim 1$ to $\sim 10^2$. The large distribution of SNR is unfavorable for the optimization of the deep learning model. Furthermore, compared to the SNR settings in the work of Ruan et al. \cite{ruan2024premerger}, we aim to further improve the model's performance at lower SNR values. Based on the above considerations, we set the SNR of the training set between 10 and 50, with each data point representing a 30-day whitened inspiral signal. The data has a sampling frequency of 0.1 Hz, so the length of each data sample is 259,200. In accordance with the aforementioned settings, we generate a total of 50,000 training samples.

\section{\label{sec:method}Methodology}

\begin{figure*}[hbt]
\includegraphics[width=0.98\textwidth]{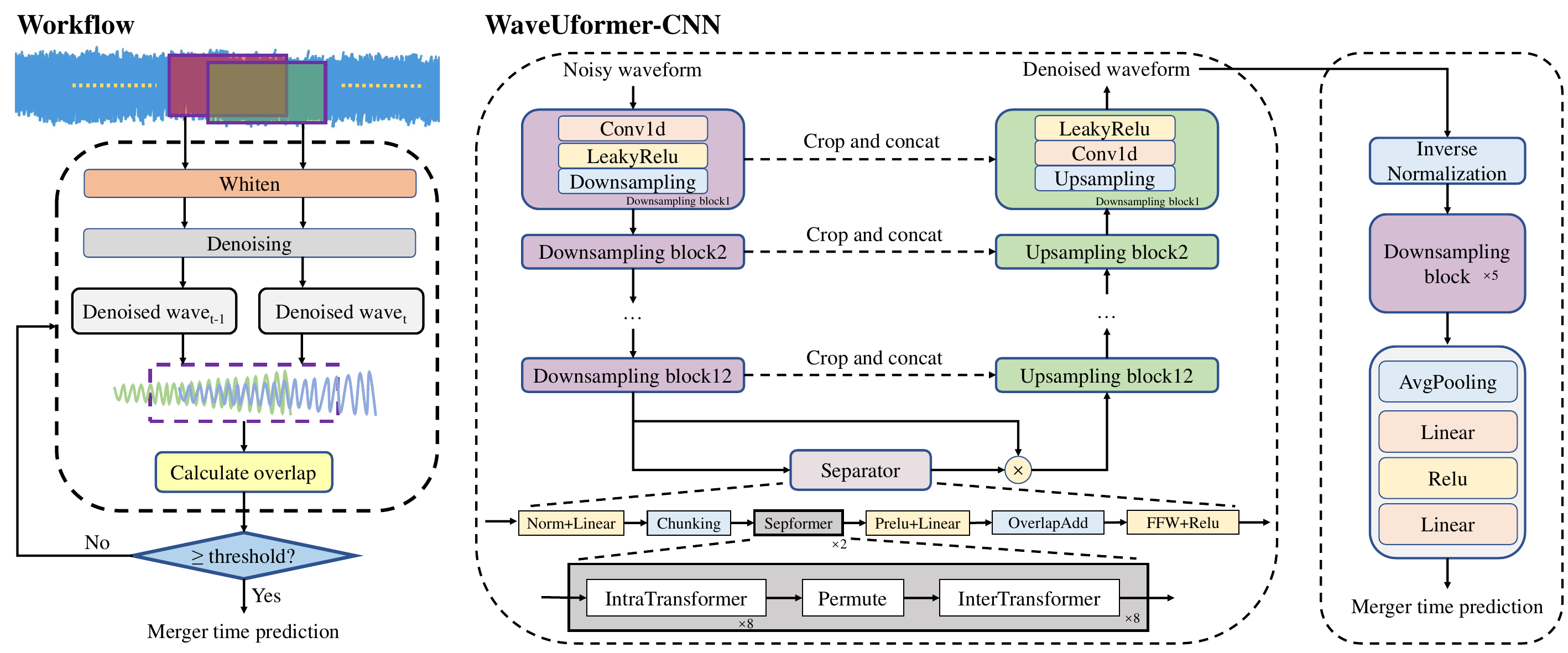}
\caption{\label{fig1}On the left, we present our workflow, where the data undergoes whitening followed by denoising. We compute the overlap between the denoised data from two adjacent time windows to determine whether it exceeds a predefined threshold, which is used to assess whether a GW signal has been detected. Once a signal is detected, we predict its corresponding merger time. The core structure of our workflow, WaveUformer-CNN, is shown on the right. WaveUformer is used for denoising the signal, while the CNN is responsible for predicting the merger time of the denoised signal.}
\end{figure*}

\begin{figure*}[hbt]
\includegraphics[width=0.98\textwidth]{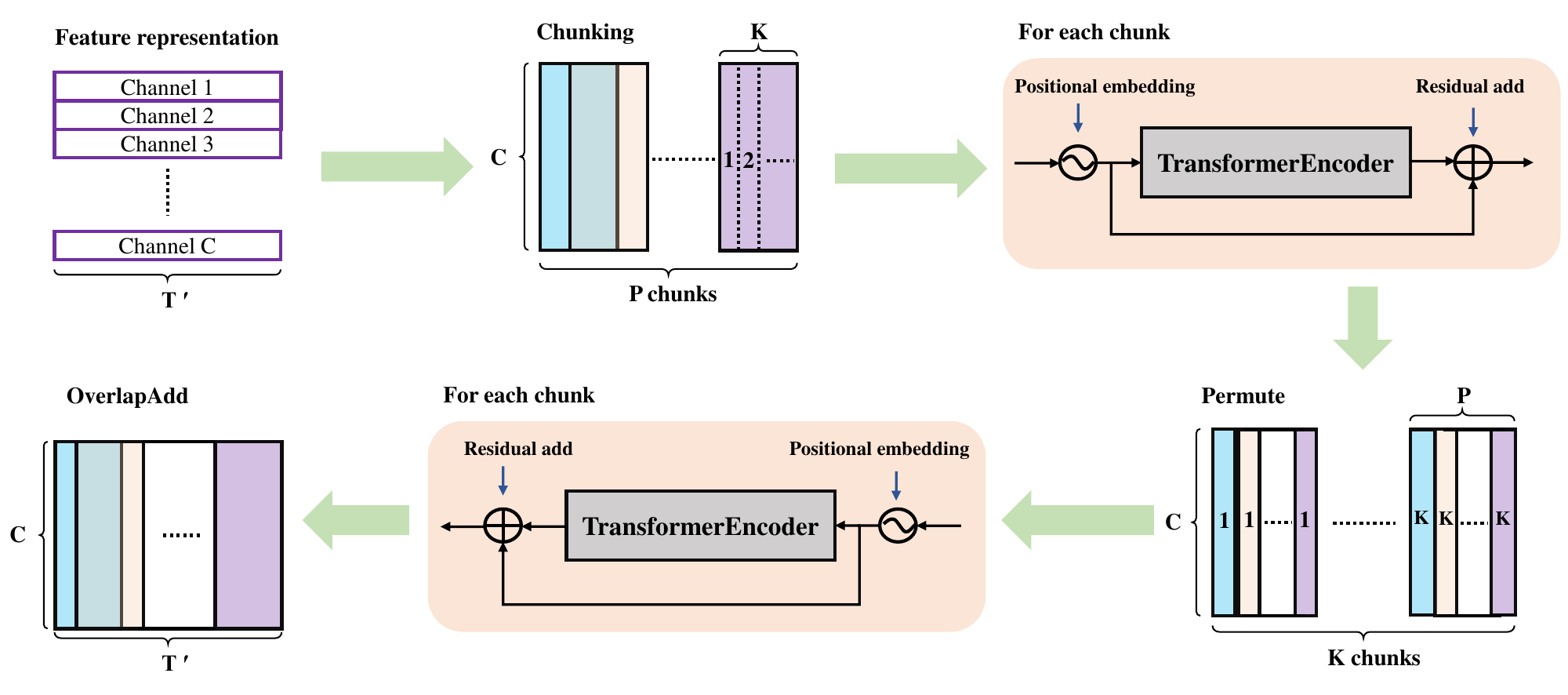}
\caption{\label{sepformer}This panel illustrates how we extract the complex GW feature mask from the downsampled feature representation of the data. We divide the data into blocks and permute them to obtain representations of short-term and long-term patterns. For these two distinct pattern representations, we use a transformer for feature extraction, ultimately obtaining the GW feature mask.}
\end{figure*}

In this work, we designed a workflow as shown on the left side of Fig~\ref{fig1} for denoising, detection, and merger time prediction of inspiral signals. This is a deep learning-driven workflow, and the detailed architecture of the deep learning model is presented on the right side of Fig~\ref{fig1}. It is worth noting that the event rate of MBHB overlap detections by the Taiji detector is currently under investigation. Therefore, this work assumes a scenario where MBHBs are relatively far apart, which we believe serves as a good starting point for further research.

Our workflow employs a 30-day time window with a 1-hour moving step to process the GW data stream. Data within the current time window is first subjected to whitening processes, then fed into a denoising model to obtain the denoised waveform. Given the moving step, there exists an overlap of 29 days and 23 hours between the current and the previous time windows. We compute the overlap within this overlapping window between the denoised waveforms of the current and previous windows. If the overlap exceeds a predefined threshold, it indicates the detection of a GW signal. The efficacy of this approach is based on the capability of a well-trained denoising deep learning network to effectively extract deep patterns present in GW sequences. This results in the output of similar denoised waveforms and high overlap in the overlapping window of consecutive time windows. In contrast, such patterns are absent in pure noise signals, leading the model's outputs to tend towards randomness, thereby resulting in low overlap in the overlapping windows. Upon detection of a GW signal, the denoised GW data is then fed into a prediction network to estimate the merger time.

Our denoising model is based on the WaveUNet \cite{stoller2018wave} architecture, with Sepformer \cite{subakan2021attention} employed in the deeper feature space to enhance the model's expressive power. The input data \( x \in \mathbb{R}^T \), where \( T \) represents the length of the data, is first processed through 12 downsampling blocks, each consisting of 1D convolution, activation functions, and downsampling operations, to progressively extract deep features, resulting in a deep feature matrix \( m \in \mathbb{R}^{C \times T'} \), where \( C \) represents the channel dimension of the final data, and \( T' \) denotes the feature length of each channel. This feature matrix is then passed through the separator to obtain a GW feature mask \( mask \in \mathbb{R}^{C \times T'} \), which captures the latent GW features within the deep feature matrix. We multiply the deep feature matrix by the GW feature mask to obtain the separated GW feature matrix \( \hat{m} \in \mathbb{R}^{C \times T'} \). During the upsampling process, 12 upsampling blocks, each consisting of 1D convolution, activation functions, and upsampling operations, are employed to progressively reconstruct the data. Notably, during each upsampling block, the upsampled feature matrix is cropped and concatenated with the corresponding downsampled feature matrix, which allows the model to effectively utilize both shallow and deep features. This structure greatly reduces the information loss incurred during downsampling. Finally, after the last upsampling block, we obtain the denoised data \( \hat{x} \in \mathbb{R}^T \).

In our denoising network, the core component is the separator module, which is also a transformer-based structure. The transformer is a widely proven effective architecture for capturing various complex patterns present in the data \cite{ruan2024premerger,wang2024waveformer,subakan2021attention,vaswani2017attention,kirillov2023segment,dosovitskiy2020image}. The time complexity of the transformer is generally considered to scale quadratically with the length of the data \cite{vaswani2017attention}. In this work, the transformer block is applied to the downsampled deep feature matrix, effectively enhancing the model's expressive power without significantly increasing the model's time complexity. In this separator, \( m \) first passes through a layer normalization (LN) \cite{ba2016layer} followed by a linear layer. Subsequently, as shown in Fig~\ref{sepformer}, the feature map is divided into \( P \) overlapping sub-blocks of size \( C \times K \). For each chunk, we employ an IntraTransformer to learn the short-term dependencies present within the chunk. Subsequently, the output block data undergoes permutation, where the size of each new block becomes \( C \times P \). The data within each new block represents an integration of the same-position data from all previous blocks, thus each new block contains global information. These new blocks are then fed into an InterTransformer to learn the long-term dependencies within the chunks. Finally, we apply overlap-add \cite{luo2020dual} to restore the feature representation to the size of \( C \times T' \). The recovered feature representation is then passed through a feed-forward network (FFN) followed by a ReLU activation function, ultimately producing the \( mask \). It is worth noting that both the IntraTransformer and InterTransformer employ the same structure as the standard Transformer. The names are used solely to distinguish their respective roles in extracting short-term dependencies and long-term dependencies within the sequence.

We denote the input to the Transformer as \( z \). Initially, a sinusoidal positional encoding \( e \) is added to the input \( z \), resulting in:
\begin{equation}
z' = z + e \,.
\end{equation}
Since the Transformer operates in parallel, it inherently loses the sequential information present in the data. Positional encoding is used to provide the model with a sense of order by encoding the positional information of the sequence and adding it to the sequence. For specific implementation details of the positional encoding, please refer to \cite{vaswani2017attention}. Subsequently, multiple Transformer layers are applied. Within each Transformer layer \( g(\cdot) \), layer normalization (LN) is first applied, followed by multi-head attention (MHA), which can be represented as follows:
\begin{equation}
z'' = \text{MHA}(\text{LN}(z')) \,.
\end{equation}
Each attention head computes scaled dot-product attention among all elements in the sequence. The data then passes through a FFN, which can be expressed as follows:
\begin{equation}
z''' = \text{FFN}(\text{LN}(z'' + z')) + z'' + z \,.
\end{equation}
Thus, the entire Transformer block \( f(\cdot) \) is defined as:
\begin{equation}
f(z) = g(z + e) + z \,.
\end{equation}
Our denoising model incorporates two Sepformer blocks. Within each Sepformer block, there are eight IntraTransformer blocks and eight InterTransformer blocks.

For the merger time prediction model, the input data is the denoised waveform, which first undergoes inverse normalization. We reuse the downsampling blocks from the denoising network to extract deep feature representations of the denoised waveform. These features are then processed through an average pooling layer followed by two linear layers, ultimately yielding the predicted merger time.

\section{\label{sec:strategy}Training strategy}

In this work, we train both the denoising and prediction models. For the denoising model, the input data consists of noisy data with injected signals, while the template waveforms serve as the labels. For the prediction model, the input data is the denoised output, with the time intervals to the merger event used as the labels. For both models, we employ the Mean Squared Error (MSE) loss function, defined as:
\begin{equation}
\text{MSE} = \frac{1}{n} \sum_{i=1}^{n} (\text{y}_{\text{pred},i} - \text{y}_{\text{true},i})^2 \,,
\end{equation}
where \( \text{y}_{\text{pred},i} \) represents the predicted values, \( \text{y}_{\text{true},i} \) represents the true values, and \( n \) is the number of samples in the dataset. Specifically, for the denoising model, \(\text{y}_{\text{pred},i}\) represents the denoised waveform, while \(\text{y}_{\text{true},i}\) represents the template waveform. For the merger time prediction model, \(\text{y}_{\text{pred},i}\) denotes the predicted merger time, and \(\text{y}_{\text{true},i}\) corresponds to the actual merger time.

We choose the Adam optimizer \cite{kingma2014adam} combined with a warmup \cite{he2016deep} and cosine annealing \cite{loshchilov2016sgdr} learning rate schedule to optimize our models effectively. The maximum learning rate is set to \(1 \times 10^{-3}\), and the minimum to \(1 \times 10^{-5}\). The total training spans 100 epochs, with the first three epochs constituting the warmup period, during which the learning rate increases from \(1 \times 10^{-5}\) to \(1 \times 10^{-3}\). This warmup ensures stability in the deeper layers of the training. Afterward, the learning rate enters a cosine annealing phase, gradually decreasing back to \(1 \times 10^{-5}\). This scheduling provides a smooth optimization space search strategy. In the early stages of cosine annealing, the learning rate is typically set to a large value, which helps in rapidly exploring the optimization space. Later, the learning rate is systematically reduced according to the cosine schedule, leading to a smoother optimization trajectory. Compared to traditional optimization techniques, cosine annealing enables a more comprehensive exploration of the optimization space, thereby helping to escape local optima and aiding in optimal convergence. All models are implemented using the Pytorch \cite{paszke2019pytorch} framework and were trained 10 hours on three NVIDIA A100 40GB GPUs.

\section{\label{sec:results}Results}

The chirp mass significantly influences the duration and amplitude variation of the GW signals during the inspiral phase. Therefore, our test set includes three different chirp masses: $1 \times 10^5 M_{\odot}$, $1 \times 10^6 M_{\odot}$, and $1 \times 10^7 M_{\odot}$. All other parameters will remain consistent with those listed in Tab~\ref{tab:data}. We will evaluate the denoising performance and predictive accuracy of our model under these conditions.

\subsection{\label{sec:denoised}Signal denoising}

In this test, the SNR for each dataset, corresponding to a different chirp mass, is uniformly sampled from 10 to 50, with each dataset comprising 5000 samples. All data are processed by our denoising model to obtain the corresponding denoised outputs. We evaluate the recovery of phase and amplitude in the denoised data. Phase recovery is assessed by calculating the overlap between the denoised data and the template data, while amplitude recovery is evaluated by comparing the SNR of the denoised data with that of the template data.

\begin{figure*}[htb]
\centering
\includegraphics[width=0.98\textwidth]{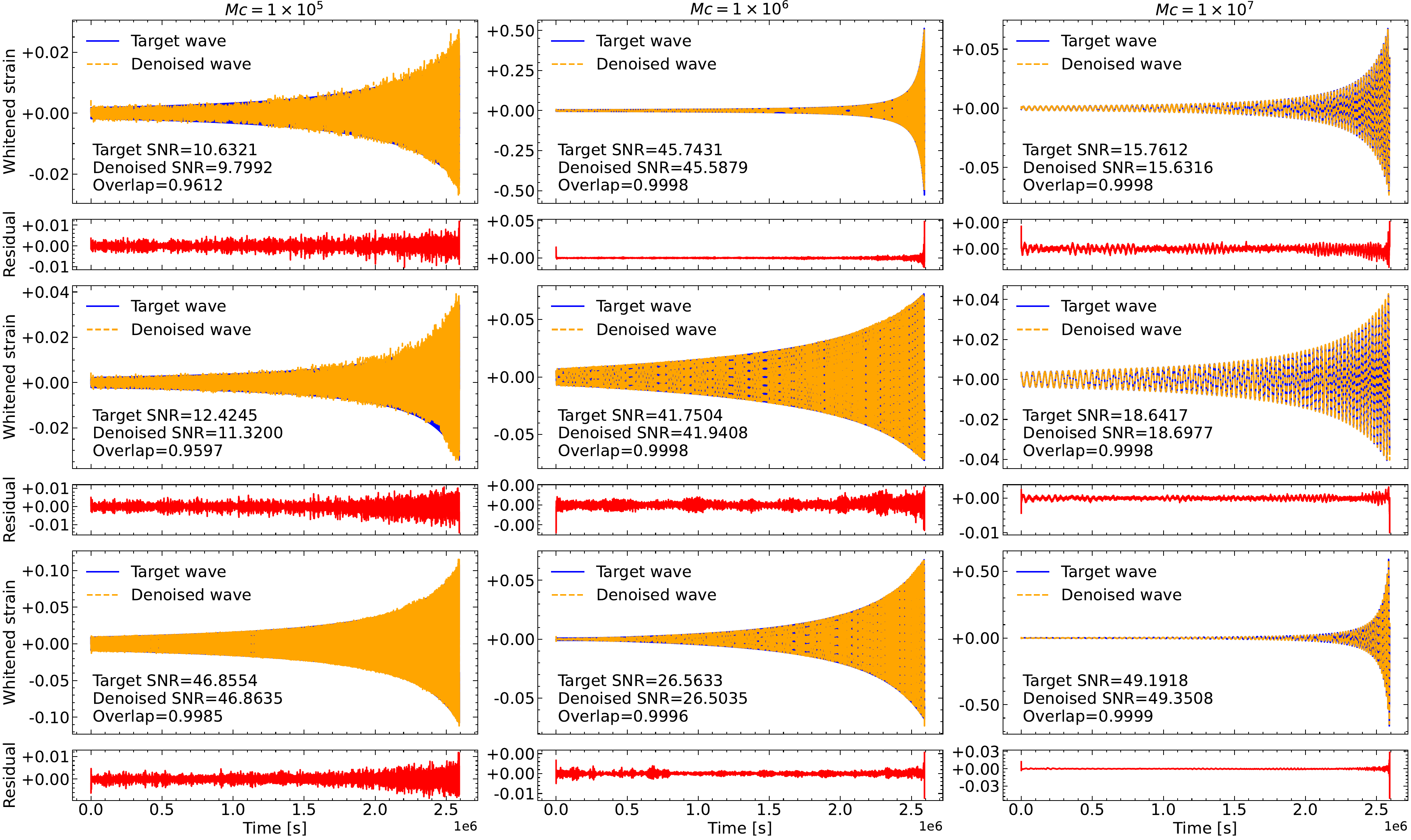}
\caption{\label{fig:denoise_1}From left to right, each column of panels displays the model extraction results for chirp masses of \( 1 \times 10^5 M_{\odot} \), \( 1 \times 10^6 M_{\odot} \), and \( 1 \times 10^7 M_{\odot} \), respectively. Below each panel, we present the residuals between the denoised data and the template data.}
\end{figure*}

\begin{figure*}[htb]
\centering
\includegraphics[width=0.98\textwidth]{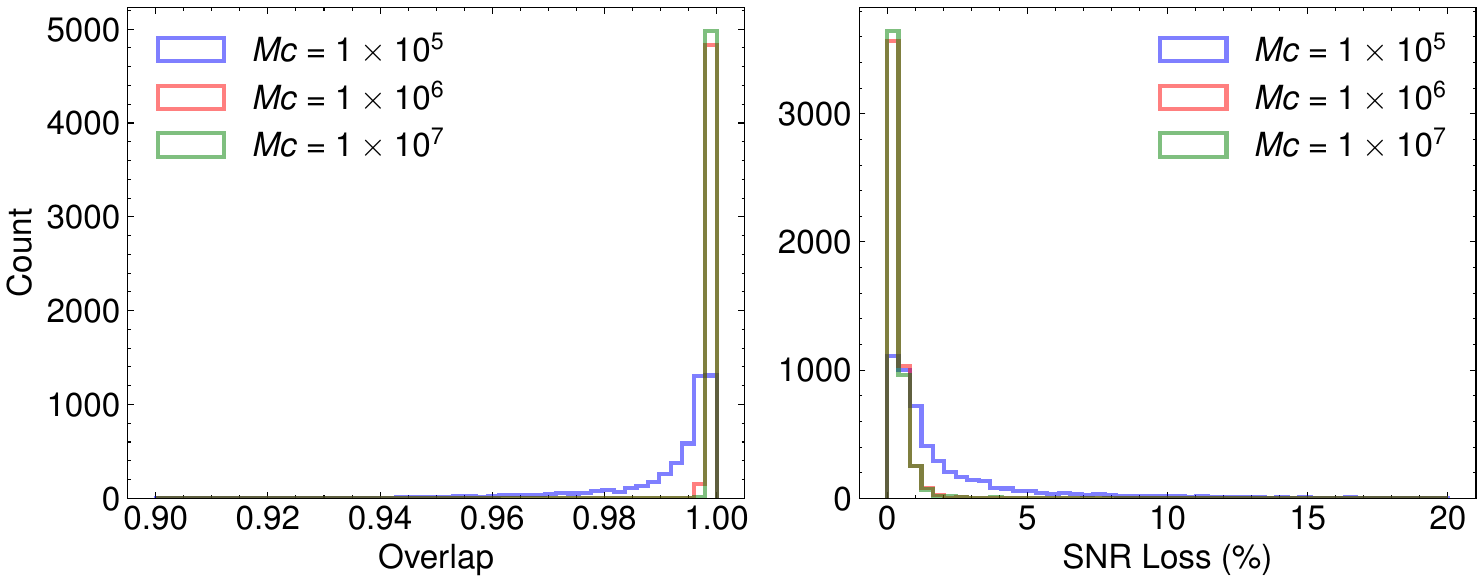}
\caption{\label{fig:denoise_2}The left panel shows the overlap distribution between the denoised data and the template data for different chirp masses. The right panel displays the percentage distribution of the SNR loss of the denoised data relative to the SNR of the corresponding template data for different chirp masses.}
\end{figure*}

Fig~\ref{fig:denoise_1} presents the denoising results for different chirp masses. It can be observed that our model demonstrates considerable denoising performance. The distribution of the results is shown in Fig~\ref{fig:denoise_2}. Since 99\% of the denoised data achieve an overlap $\geq$ 0.9 with the template data, we have set the displayed overlap range between 0.9 and 1 for ease of visualization. The results for the dataset with a chirp mass of \(1 \times 10^5 M_{\odot}\) are slightly inferior to those for other datasets. This phenomenon occurs because, for smaller chirp masses, the pre-merger inspiral and approach processes are slower, resulting in less pronounced amplitude variations and a less steep frequency variation across the entire data segment. This substantially confuses the model's judgment.

In Tab~\ref{tab:denoise}, we present detailed statistical results. It can be observed that our model precisely recovers the phase for data across different chirp masses. Overall, 91.98\% of the data, post-denoising, achieves an overlap $\geq$ 0.99 with the template data, demonstrating our model's sensitivity to phase recovery. Regarding amplitude recovery, while performance is slightly lower for datasets with small chirp mass, the overall results indicate good amplitude recovery, with 95.82\% of the denoised data has an SNR loss $\leq$ 5\% compared to the template SNR.

\begin{table}[htb]
\caption{\label{tab:denoise}The mean and standard deviation of the overlap and SNR differences between the denoised and template data.}
\begin{ruledtabular}
\renewcommand{\arraystretch}{1.2}
\begin{tabular}{lcc}
Chirp mass & Overlap & SNR loss (\%) \\
\colrule
$1 \times 10^5 M_{\odot}$ & $0.991 \pm 0.022$ & $2.807 \pm 6.976$ \\
$1 \times 10^6 M_{\odot}$ & $0.999 \pm 0.008$ & $0.429 \pm 2.458$ \\
$1 \times 10^7 M_{\odot}$ & $0.999 \pm 0.003$ & $0.375 \pm 0.917$ \\
\end{tabular}
\end{ruledtabular}
\end{table}

\subsection{\label{sec:detection}Signal detection}

\begin{figure*}[hbt]
\centering
\includegraphics[width=0.8\textwidth]{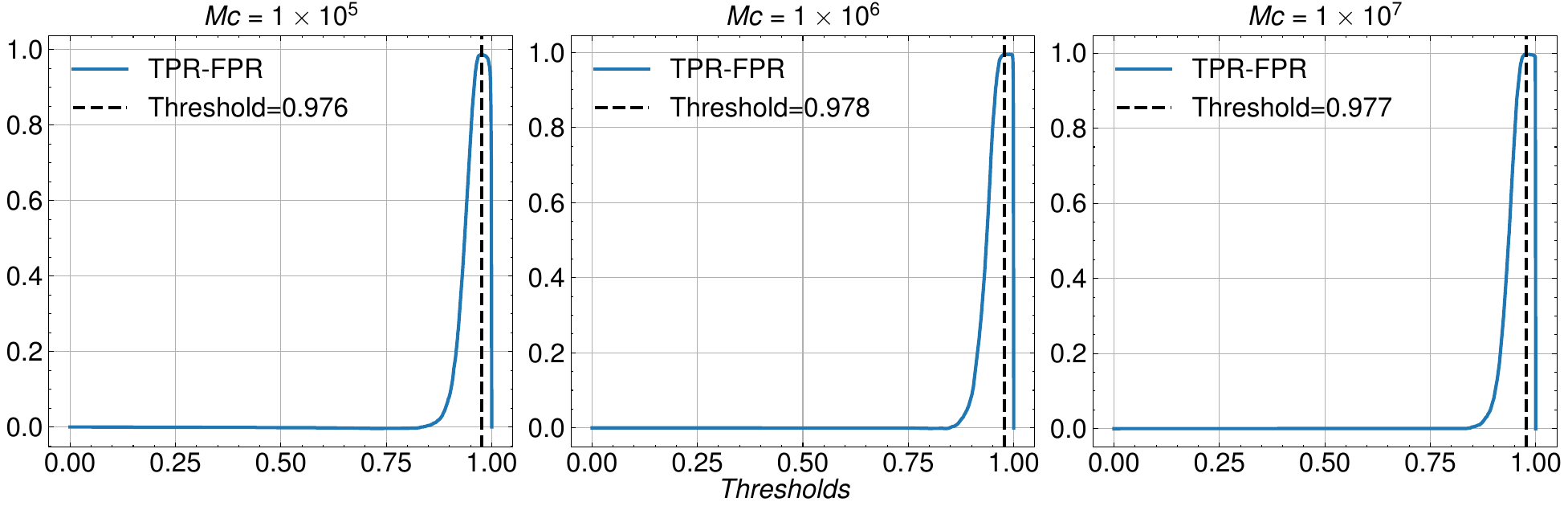}
\caption{\label{fig:cla_1}From left to right, the panels display the TPR-FPR at different thresholds for chirp masses of \(1 \times 10^5 M_{\odot}\), \(1 \times 10^6 M_{\odot}\), and \(1 \times 10^7 M_{\odot}\), respectively.}
\end{figure*}

\begin{figure*}[hbt]
\centering
\includegraphics[width=0.98\textwidth]{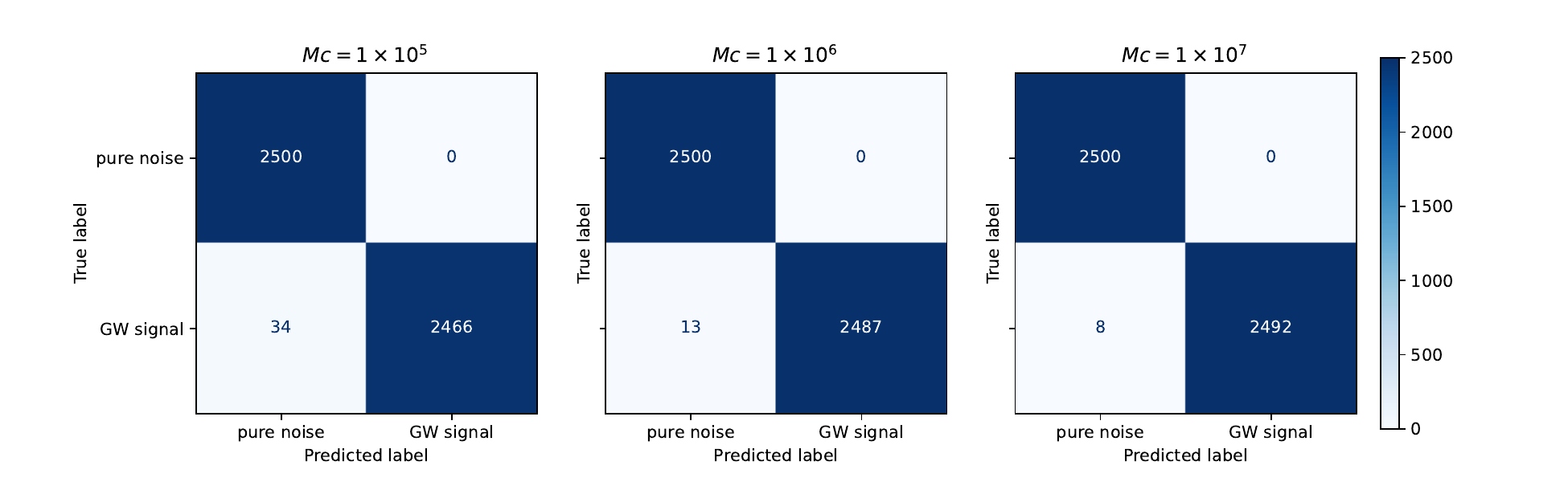}
\caption{\label{fig:cla_2}From left to right, the panels display the confusion matrices for the classification of pure noise signals and GW signals for chirp masses of \(1 \times 10^5 M_{\odot}\), \(1 \times 10^6 M_{\odot}\), and \(1 \times 10^7 M_{\odot}\), respectively. The data on the main diagonal represents the number of correct predictions within that class, while the data on the off-diagonal indicates the number of incorrect predictions for that class.}
\end{figure*}

In this test, we will generate 5,000 data pairs for each dataset corresponding to different chirp masses. These pairs consist of current moment data and data from one hour earlier. The SNR for these data pairs is uniformly sampled between 10 and 50. Each dataset is a balanced binary classification dataset, with half of the samples being GW signals and the other half being pure noise. According to the design of our workflow, we calculated the overlaps for the data pairs in each dataset. To explore the selection of a reasonable threshold, we calculated the true positive rate (TPR) and false positive rate (FPR) for each dataset, as shown in Fig~\ref{fig:cla_1}. We selected 0.978 as the threshold. At this value, the FPR for all datasets is zero, and the TPR is close to 1. This indicates that our model exhibits high sensitivity to GW signals at this threshold while not responding to pure noise. In Fig~\ref{fig:cla_2}, we present the confusion matrices for each dataset at the selected threshold. Even for a chirp mass of $1 \times 10^5 M_{\odot}$, only 34 GW signals were missed, resulting in a detection accuracy of 98.64\%. Furthermore, the detection accuracy increases with higher chirp masses. These results demonstrate the robustness and rationality of our workflow.

\subsection{\label{sec:detection}Merger time prediction}

\begin{figure*}[htb]
\centering
\includegraphics[width=0.98\textwidth]{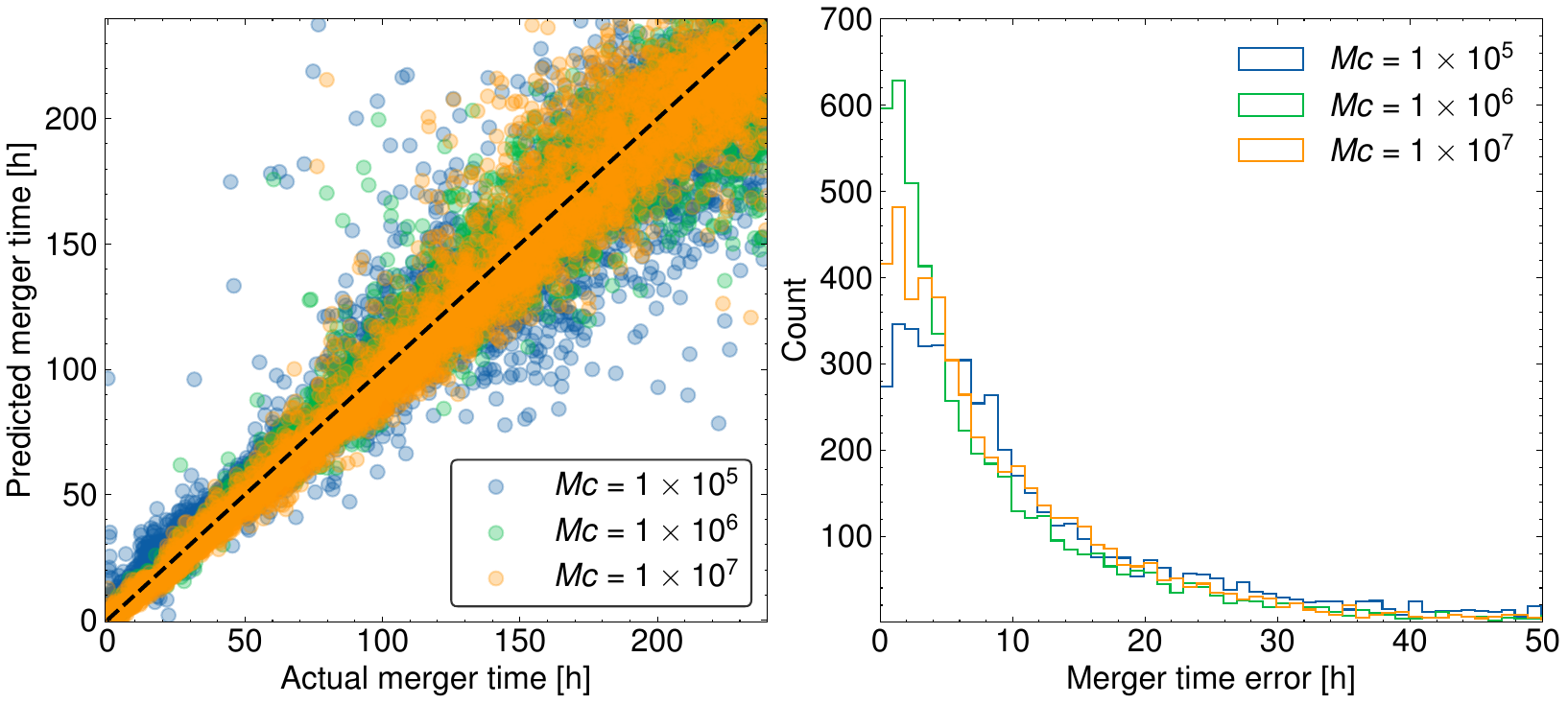}
\caption{\label{fig:pred_1}The left panel shows the distribution of merger time predictions for chirp masses of \(1 \times 10^5 M_{\odot}\), \(1 \times 10^6 M_{\odot}\), and \(1 \times 10^7 M_{\odot}\). The x-axis represents the actual time until the merger occurs, and the y-axis represents the predicted time until the merger. Points closer to the black dashed line in the diagram indicate smaller prediction errors. The right panel presents histograms of merger time errors for chirp masses of \(1 \times 10^5 M_{\odot}\), \(1 \times 10^6 M_{\odot}\), and \(1 \times 10^7 M_{\odot}\).}
\end{figure*}

In this test, we generated datasets with chirp masses of \(1 \times 10^5 M_{\odot}\), \(1 \times 10^6 M_{\odot}\), and \(1 \times 10^7 M_{\odot}\), each containing 5000 samples. The SNR within each dataset was uniformly sampled between 10 and 50. The generated data was first input into the denoising network to obtain the denoised waveforms, which then served as inputs to the merger time prediction network, yielding predicted merger times. The distribution of results for each dataset is depicted in Fig~\ref{fig:pred_1}(left), where a general trend is observed: the prediction error decreases as the merger event approaches. This is attributed to the increase in signal strength near the merger, which enhances the recovery of amplitude and phase in the denoised signal, thereby leading to more accurate predictions. Fig~\ref{fig:pred_1}(right) illustrates the histogram of prediction errors, revealing that signals with a chirp mass of \(1 \times 10^5 M_{\odot}\) exhibit larger prediction errors compared to the other two classes. This suggests that the smaller chirp mass data has less pronounced amplitude variations, making denoising more difficult and amplifying the prediction error for the merger time.

To further explore the overall performance of our model, we conducted a statistical analysis of the results across the entire test set based on the time distance to the merger event. The results, shown in Tab~\ref{tab:prediction}, indicate that the average prediction error decreases and the distribution becomes more stable as the dataset approaches closer to the merger time. The time to merger of our dataset ranges from 0 days to 10 days, with 89.1\% of the data achieving a absolute prediction error of $\leq$ 24 hours. This demonstrates that our model is capable of making reasonably accurate predictions of merger times well in advance of the event.

\begin{table}[htb]
\caption{\label{tab:prediction}The mean and standard deviation of the prediction error between the denoised and template data.}
\begin{ruledtabular}
\renewcommand{\arraystretch}{1.2}
\begin{tabular}{lc}
Time to merger & Prediction error \\
\colrule
$\leq$ 10 days & $10.65 \pm 12.57$ hours\\
$\leq$ 7 days & $7.96 \pm 10.23$ hours\\
$\leq$ 5 days & $6.01 \pm 8.58$ hours\\
$\leq$ 3 days & $4.37 \pm 6.07$ hours\\
$\leq$ 1 day & $3.42 \pm 4.42$ hours\\
\end{tabular}
\end{ruledtabular}
\end{table}

\section{\label{sec:summary}Concluding remarks}

In this work, we developed an efficient deep learning model designed for denoising data during the inspiral phase of MBHB and predicting the merger time. Our model consists of two main components: a denoising structure and a prediction structure. This design is motivated by the fact that signals during the inspiral phase are often deeply buried in overwhelming noise, and their low SNR makes them unsuitable for direct prediction of merger time. Through our denoising model, faint GW signals can be extracted, enabling accurate predictions of merger times. Such a structural design empowers our model to detect signals with extremely low SNRs.

Advance warnings for MBHB mergers are valuable for possible multi-messenger observations with EM and GW detections. However, the SNR of MBHB signals accumulates in a highly nonlinear fashion, with the final hour before merger contributing as much as 99\% of the total SNR \cite{chen2024near}. As a result, the inspiral segment exhibits a significantly lower SNR, which makes detection challenging. This challenge arises mainly from the need to process long-duration signals, which introduce significant complexity into model training. Short data windows capture detectable features only near the merger phase, significantly limiting early warning capabilities. To address this, we integrate data dimensionality reduction and feature separation techniques to denoise signals up to 30 days long, allowing sufficient SNR to accumulate before the merger. This enalbles efficient detection during the inspiral phase of MBHBs. Our work also has implications for detecting other faint GW signals. Additionally, our model predicts merger times based on denoised waveforms, facilitating more precise planning for multi-messenger observations of MBHB’s EM and GW signals from its merger phase.

We evaluated our model in three key scenarios: signal denoising, signal detection, and merger time prediction. For signal denoising, our model effectively restored the amplitude and phase of noisy signals, achieving an overlap $\geq$ 0.99 with the signal templates for 91.98\% of the denoised samples, and an SNR loss $\leq$ 5\% for 95.82\% of the samples, demonstrating the effectiveness of our denoising network. For signal detection, we designed a workflow to differentiate between pure noise and GW signals. We explored optimal threshold settings to avoid false responses to noise while maintaining high sensitivity to GW signals, achieving a detection accuracy exceeding 98.64\% on our test dataset. In terms of merger time prediction, our model demonstrated considerable prediction accuracy; for data occurring no more than 10 days before the merger, the absolute prediction error generally did not exceed 24 hours.

In the future, we aim to enhance the model's capability to process data with low chirp masses, including improvements in training strategies and architectural design. Furthermore, in this work, the current method is only applicable to the detection of a single inspiral signal. We are currently conducting research on the event rate of multiple inspiral signal overlaps detected by the Taiji detector. Once we have reached definitive conclusions, we plan to extend the research to the detection of multiple inspiral signal overlaps. We believe this work will serve as a solid starting point. We also plan to test the robustness of our model confront with possible data anomalies in future works. Space-based GW detection missions may encounter data anomalies and non-stationarities, such as data gaps caused by routine maintenance or hardware issues \cite{dey2021effect,xu2024gravitational}, or acceleration glitches in the GRS system \cite{armano2022transient}. Moreover, the waveform model used in this study, IMRphenomD, does not include higher-order modes, and we leave these more realistic challenges for future exploration.

\begin{acknowledgments}
This work is supported by the National Key Research and Development Program of China under Grant Nos. 2021YFC2201901, 2021YFC2201903, 2020YFC2200601, 2020YFC2200901, and 2021YFC2203004, as well as the National Science Foundation of China (NSFC) under Grant Nos. 12405076, 12347103, and 12247187. Additionally, this work is also supported by the International Partnership Program of Chinese Academy of Sciences, No. 025GJHZ2023106GC. We thank the anonymous referee for valuable comments and suggestions that helped us to improve the manuscript.
\end{acknowledgments}

\bibliography{apssamp}

\end{document}